# A synchrotron-based kilowatt-level radiation source for EUV lithography

Bocheng Jiang [1,2], Chao Feng [1,2*], Changliang Li [1], Zhenghe Bai [3], Weishi Wan [4], Dao Xiang [5], Qiang Gu [1,2], Kun Wang [1,2,6], Qinglei Zhang [1], Dazhang Huang [1,2], Senyu Chen [7]

[1] Shanghai Advanced Research Institute, Chinese Academy of Sciences,
Shanghai 201204, China

[2] Shanghai Institute of Applied Physics, Chinese Academy of Sciences,
Shanghai 201800, China

[3] National Synchrotron Radiation Laboratory, USTC,
Hefei 230029, China

[4] School of Physical Science and Technology, ShanghaiTech University, Shanghai 201210,
China

[5] Key Laboratory for Laser Plasmas (Ministry of Education), School of Physics and
Astronomy, Shanghai Jiao Tong University, Shanghai 200240, China

[6] University of the Chinese Academy of Sciences,
Beijing 100049, China

[7] Institute of High energy physics, Chinese Academy of Sciences, Beijing 100049, China

A compact damping ring with limited circumference of about 160 m is proposed for producing kilowatt-level coherent EUV radiation. The electron bunch in the ring is modulated by a 257nm wavelength laser with the help of the angular dispersion induced micro-bunching method [C. Feng and Z. Zhao, Sci. Rep. 7, 4724 (2017)]. Coherent radiation at 13.5 nm with an average power of about 2.5 kW can be achieved with the state-of-the-art accelerator and laser technologies.

I. Introduction

Radiation from accelerator based light sources for optical lithography had been studied for a long time [1, 2]. Accelerator based light sources for lithography gets several advantages. It is a clean light source without debris contaminating the optics and it is convenient tuning the wavelength without a major technical change. It has been confirmed by the semiconductor industry that 13.5nm wavelength extreme ultraviolet (EUV) lithography will be the route for edge wafer manufacturing. High power EUV light source is one of the key technologies for EUV lithography. The EUV source of average power beyond 500W is the cutting edge of the research both for laser-produced plasma (LPP) light sources and accelerator based light sources.

Yet, the average power of the spontaneous EUV radiation from an electron storage ring is only several watts even with extremely high beam current and long undulators. Using micro-bunched electron beams is currently the most effective way to enhance the average power of accelerator based light source since the output power is proportional to square of the number of electrons in the micro-bunches [3, 4]. For storage rings, the leading concept for realizing this kind of light source is the steady-state micro-bunching (SSMB) [5-7]. Currently, one of the critical issues of SSMB is how to further compress the micro-bunch to make it shorter than the EUV wavelength on a turn-by-turn basis.

* fengchao@zjlab.org.cn

Micro-bunches with durations at the EUV and soft X-ray wavelength scale can be achieved by utilizing the angular dispersion induced micro-bunching (ADM) technique [8], which can precisely tailor the electron beam longitudinal distribution with the aid of an optical laser. With proper setting of the modulation amplitude and the dispersion chicane, the bunching factor can be written:

$$b_n = J_n(nk_s\xi\frac{\Delta\gamma}{\gamma})e^{-\frac{1}{2}(nk_s\eta\sigma_{y'})^2}, \quad (1)$$

where $b_n$ is the bunching factor of the n[th] harmonic, $k_s$ is the wave number of the seed laser, $\xi$, $\eta$ are the momentum compaction and dispersion function of the dispersive chicane respectively, $\gamma$ is the relativistic parameter for the mean beam energy, $\Delta\gamma$ is the energy modulation amplitude induced by the seed laser. $\sigma_{y'}$ is the vertical angular divergence of the electron beam. When vertical angular divergence of the electron beam is extraordinary small, unprecedented high harmonic can be achieved

However, this manipulation processes, or so called the modulation, will increase the electron beam energy spread and the vertical emittance, resulting a limited repetition rate [9] even with a demodulation [10] that cancels most parts of the modulation. For the EUV radiation purpose, the beam energy is optimized to a few hundreds of MeV, for which the synchrotron radiation damping is very weak, the damping time is several tens or even hundreds of milliseconds. The residual perturbation caused by the modulation needs thousands of turns being damped down. A storage ring with shorter damping time is highly desired to eliminate the perturbations rapidly and to achieve a higher modulation repetition rate as well as getting higher average radiation power.

Damping rings have been widely investigated for colliders [11, 12, 13]. Damping wiggler is an indispensable device in the damping ring that reduces both the damping time and transverse emittances. Nevertheless, the vertical focusing effect of strong damping wiggler will significantly distort the linear beam optics, especially when the beam magnetic rigidity (beam energy) is low (hundreds of MeV), sometimes the periodic lattice solutions do not exist anymore [14]. In medium energy rings, superconducting wigglers (SWs) with limited length are used for both colliders and synchrotron radiation facilities [15, 16, 17]. Long SWs in medium energy storage ring will create huge radiation power, making great technical challenges for photon absorbers [18]. Worse still, damping wiggler also contributes remarkable nonlinear effects that may shrink the dynamic aperture (DA) and the momentum aperture (MA) [19], resulting in a limited lifetime of the electron beam.

In this paper, a compact EUV light source that combines the dumping ring and the ADM techniques is proposed. A special design for SWs with quadrupole poles inside is given and studied. This design splits the focusing equally between horizontal and vertical planes, making the transfer matrixes in both planes identical ones. As a result, the beta functions in the wiggler can be very small which minimizes the nonlinear effects. The MA of the damping ring is optimized to a large value and a dedicated demodulation bypass line is given to ensure a reasonable beam lifetime for high current operation. Three-dimensional simulations have been performed and the results indicate the generation of kilowatt-level EUV radiation at 13.5 nm with current available technologies.

* fengchao@zjlab.org.cn

## II. Equally focused wiggler

The wiggler magnet with wide enough poles present a longitudinal field written as [14],

$$B_z = B_0 \sin(k_p z) \sinh(k_p y) = B_0 \sin(k_p z) \left(k_p y + \frac{(k_p y)^3}{3!} + \cdots\right), \quad (2)$$

where $z$ is the longitudinal direction along beam axis. When the beam wiggles in the horizontal plane, $B_z$ will produce a vertical force. In Eq.(2), $B_z$ is proportional to $y$ for the first order approximation which acts as a quadrupole field in vertical (V) plane. While in horizontal (H) plane, the beam acts likely passing through a drift. The transfer map difference between V/H planes makes it difficult to match in the ring.

This difference can be eliminated by designing the wiggler poles as wedge magnets [14]. This method is effective when magnetic field is not so strong. For the strong wigglers such as SWs, the limited wedge angle is insufficient to balance the focus between V/H planes. Several types of planar wigglers, such as the alternate pole canting wiggler, had been proposed to produce additional horizontal focusing [20, 21]. However, these field manipulation methods are convenient for the permanent magnet wiggler. While for SWs, the magnet field is beyond saturation of the yoke, the quadrupole field quality is difficult to control by introducing gradient of the poles.

Here we propose inserting sets of quadrupoles in the wiggler to balance the transverse focuses in both planes. The schematic layout of the design is given in Fig.1. where the poles of orange color are quadrupoles. The equally focused wiggler is composed by a segment of wiggler followed by a quadupole and in repetition. This model is simulated by ELEGANT code [22] with canonical integration method. The structure is compact and effective, identical transfer matrices can be found in both planes with proper choice of the parameters as shown in Table 1.

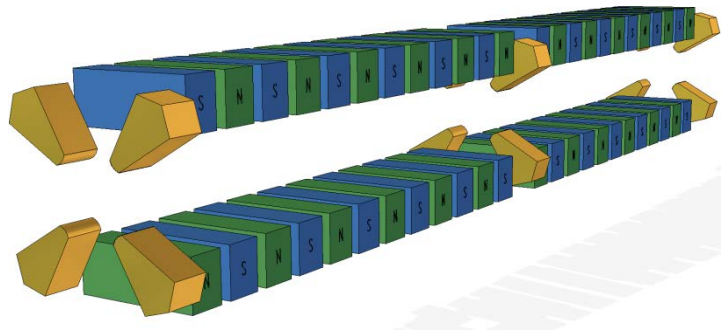

Figure 1. Schematic view of equally focused wiggler

The optimized Twiss parameters are shown in Fig 2, where the beta function is low and in periodicity. In this setting, the technical challenges had been fully considered, the wiggler is segmented to 3 sections, each with two segments 0.9m pole length sandwiched by two 0.1m and one 0.2m long quadrupoles. Two 0.4m drift space at both ends for cryogenic tank had been reserved, make sure a single wiggler to a reasonable length of 3.0m. The peak magnetic field is 5.66 Tesla which is achievable with superconducting techniques.

* fengchao@zjlab.org.cn

Table I. Beam parameters for equally focused superconducting wiggler

| Parameters | Value |
|---|---|
| Beam energy (MeV) | 1000 |
| Period length (mm) | 60 |
| Peak magnetic field (Tesla) | 5.66 |
| Quad Gradient family 1 (T/m) | 13.1 |
| Quad Gradient family 2 (T/m) | 10.3 |
| Pole gap (mm) | 10 |

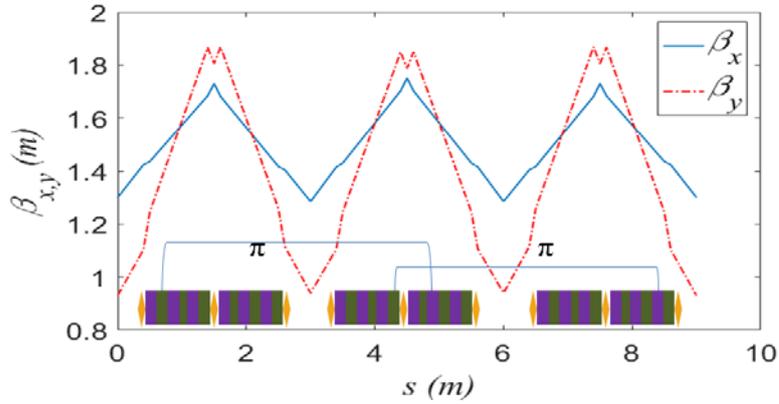

Figure 2. Twiss parameters in the wiggler.

Since the betatron phase advance of the wiggler is $2\pi$, there are many $\pi$ nodes as shown in Fig. 2, which cancels most parts of the nonlinear kicks and shapes a good nonlinear performance, as we will show in the following section.

Unlike the Robinson wiggler [23, 24], for this study the wiggler is place at the dispersion free straight section, the quadrupole fields combining with wiggler field will not redistribute the damping partition number.

It is worth to stress here that the helical undulator can also produce both horizontal and vertical focus naturally. However, by increasing the field of the helical undualtor, it will excite vertical emittance which is not compatible with the ADM as a tiny vertical emittance is highly required. This is the reason that helical undulator is not adopted in our design.

## III. Damping ring with large momentum acceptance

For high power EUV radiation from a micro-bunched electron beam, we needs beam in a storage ring with peak current more than 100A, this may result severe intra beam scattering (IBS) and Touschek effects. The relative high beam energy of 1GeV is chosen to mitigate those effects yet the energy is still suitable for EUV radiation. The Touschek lifetime strongly depends on the MA. For a low energy and high peak current ring, local momentum aperture (LMA) is the majorly consideration of the lattice design which is bounded by the nonlinear beam dynamics. LMA is usually lower in the arc where the dispersion is nonzero. The LMA will be reduced as the increase of the dispersion. The way to reduce the dispersion without rapidly rising the sextupole strength is to increase the number of the lattice cells. While considering the ring needs to be as compact as possible to get cost competitive, the number of cells is eventually chosen to be 8. There are 8 straight sections, 6 of them are accommodated by the SWs, the other 2 are for injection, extraction and RF system.

* fengchao@zjlab.org.cn

Triple-bend achromat (TBA) lattice was designed for the ring. To have a compact configuration, all bending magnets are combined-function ones. There are 3 families of chromatic sextupoles in the lattice. The two defocusing sextupoles of the same family close to the matching bending magnets have the highest integrated strength, and the horizontal betatron phase advance between these two sextupoles is about π, which is beneficial for enlarging horizontal dynamic aperture. The fractional parts of the horizontal and vertical tunes of each lattice cell are near (3/8, 5/8) for nonlinear dynamics cancellation over 8 cells.

The beam parameters with/without considering IBS effects are shown in Table II. The Twiss parameters of a half ring are shown in Fig.3 and The LMA of half ring gotten through tracking is as shown in Fig. 4.

Table II. Ring parameters

|  | W/O IBS | W/O IBS | With IBS |
|---|---|---|---|
|  | W/O SW | With SW | |
| Beam energy (MeV) | 1000 | 1000 | 1000MeV |
| Circumference (m) | 80 | 158.4 | 158.4 |
| Tune(x/y) | 11.25/5.15 | 18.27/12.17 | 18.27/12.17 |
| Horizontal emittance (nm·rad) | 3.07 | 0.42 | 1.35 |
| Energy spread | 6.63e-4 | 1.01e-3 | 1.18e-3 |
| Energy loss per turn (MeV) | 0.046 | 0.704 | 0.704 |
| Damping time(x/y/s) (ms) | 7.7/11.5/7.7 | 1.45/1.49/0.76 | 1.45/1.49/0.76 |
| RF frequency (MHz) | - | - | 499.65 |
| RF voltage (MV) | - | - | 1.2 |
| Harmonic Number | - | - | 264 |
| Bunch charge (nC) | - | - | 8.28 |
| Bunches | - | - | 190 |
| Bunch length (mm) |  |  | 9.0 |
| Beam Current (A) | - | - | 3.0 |
| Peak Current (A) | - | - | 111 |
| Betatron coupling | - | - | 0.7% |
| Touschek lifetime (hours) | - | - | 0.5 |

* fengchao@zjlab.org.cn

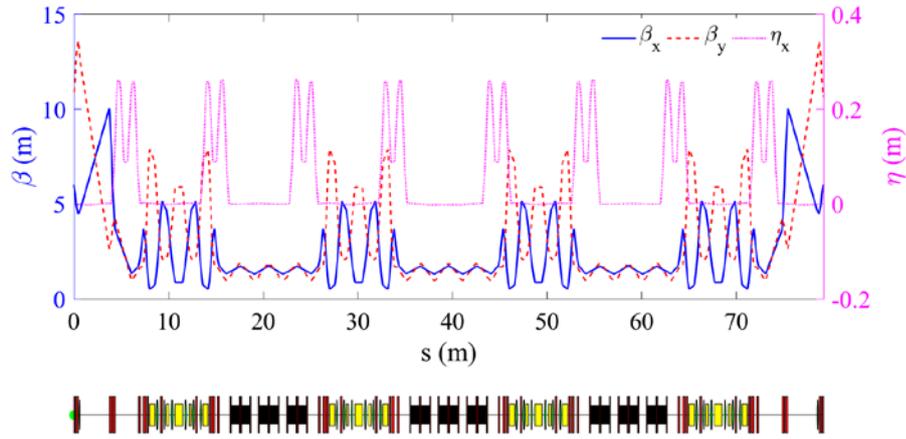

Figure 3. Twiss parameters of a half of the ring.

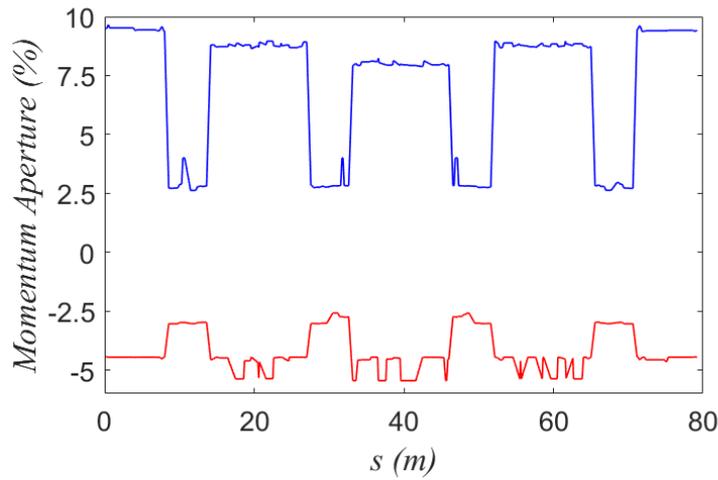

Figure 4. Local momentum aperture of half ring.

The nonlinear effects of SWs are under well control, owing to π nodes in the wiggler which cancels most of the nonlinear kick. The beta function in the wiggler is small, which minimizes the nonlinear effects of the nonlinear kick. As shown in Fig. 4, the LMA in the arc is more than 2.5%. The DA of the ring, as shown in Fig. 5, is about 10mm in horizontal plane. The main optimization target of this ring is a relative large LMA which is of great importance for the Touscheck lifetime. DA is not fully optimized, but is large enough for injection. The biggest challenge of nonlinear beam dynamics in this case is not SWs, but matching two long straight sections reserved for RF cavity and injection/extraction elements. Long straight sections break the symmetry of the ring, arousing high order driving terms deteriorate nonlinear performance.

* fengchao@zjlab.org.cn

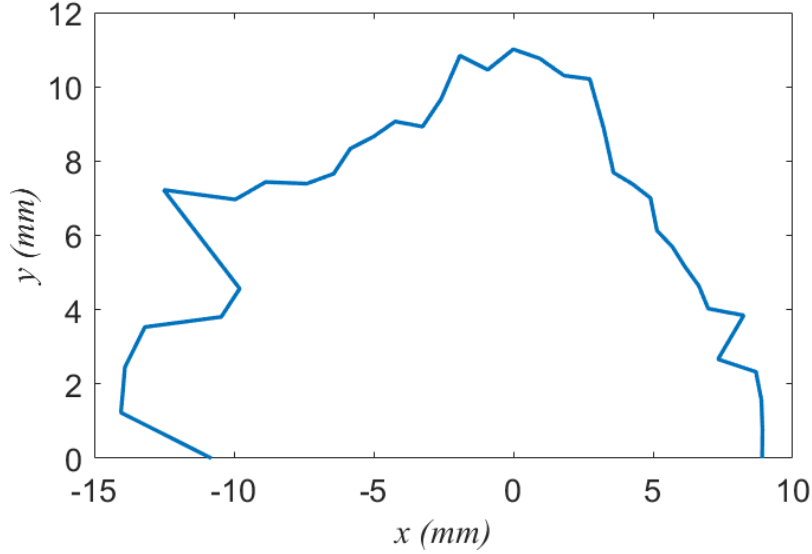

Figure 5. Dynamic aperture with superconducting wiggler

## IV. High power EUV generation

Based on the above storage ring, ADM scheme is utilized to generate micro-bunching and enhance the EUV radiation. In order to get a large bunching factor, we need an injection beam with small enough angular divergence, which means a large vertical beta function and zero alpha function[9] at the point of the vertical bend $B_0$. After the electron beam passes through the vertical bend $B_0$, the electron beam with different initial energy spread will have different angular dispersion, then the electron beam interacts with the external laser in the modulator (M). After the energy modulation, the electron beam goes through a dispersion section called dogleg, which can convert the energy modulation into the density modulation, therefore micro-bunches can be realized by properly setting the bending angle of $B_0$, the energy modulation amplitude and the dispersion of the dogleg. The main parameters for ADM are given in Table III. The micro-bunched beam can emit temporal coherent radiation through the radiator (R).

When the electron beam interacts with the laser in the modulator the energy spread will be inevitably increased. The vertical dispersion in the modulator is nonzero, which causes a vertical emittance growth simultaneously. For high power radiation purpose, we need to improve the repetition rate of the coherent radiation, thus the demodulation (D-M) of the electron beam is necessary to erase the energy modulation as to perturb the electron beam as less as possible.

As shown in Fig. 6, the M and D-M beam line is designed. The beam line gets five quadrupoles in the center with two vertical bends at both sides forms a double bend archromatic (DBA)-like structure. The $R_{56}$ generated by the doglegs is cancelled by the DBA structure, so that the $R_{56}$ between M and D-M is zero. That is to say, the beam line between M and D-M will be isochronous, under which condition the demodulation is the most effective. The sketch of damping ring and bypass beam line is as shown in Fig.7.

* fengchao@zjlab.org.cn

The beam line is symmetric, which has many advantages. It ensures that the transport line is achromatic. At the same time, the symmetrical structure can return the beam orbit to the original horizontal plane and is better for cancelling nonlinear high-order terms.

Table III. Parameters for ADM

| | |
|---|---|
| Bending angle of $B_0$(mrad) | 9.5 |
| Length of $B_0$ (m) | 0.3 |
| Laser wave length (nm) | 257 |
| Energy modulation amplitude ($\sigma_{E0}$) | 0.6 |
| R56 of dogleg | -6.15e-5 |
| Dispersion of dogleg (mm) | 6.5 |
| Distance between two bends in dogleg(m) | 0.265 |

For the modulation at 257 nm, the longitudinal position misplacement of electrons between M and D-M should be within few nm, so we need to add some sextupole magnets to correct the high-order terms. Four families of sextupoles (eight in total) are added to correct the high-order term, as shown in Fig.6. In order not to affect nonlinearity of the storage ring as well as less burden of linear optics matching, we adopt a bypass line for the beam manipulation and EUV generation, as shown in Fig. 7. The major part of the bypass beam line consists of three sections: the *core* section between M and D-M is isochronous and with controllable high order terms; the *dispersion match* section makes the whole beam line archromat in vertical plane; the *Twiss match* section matches the Twiss parameters to the rest part of the beam line.

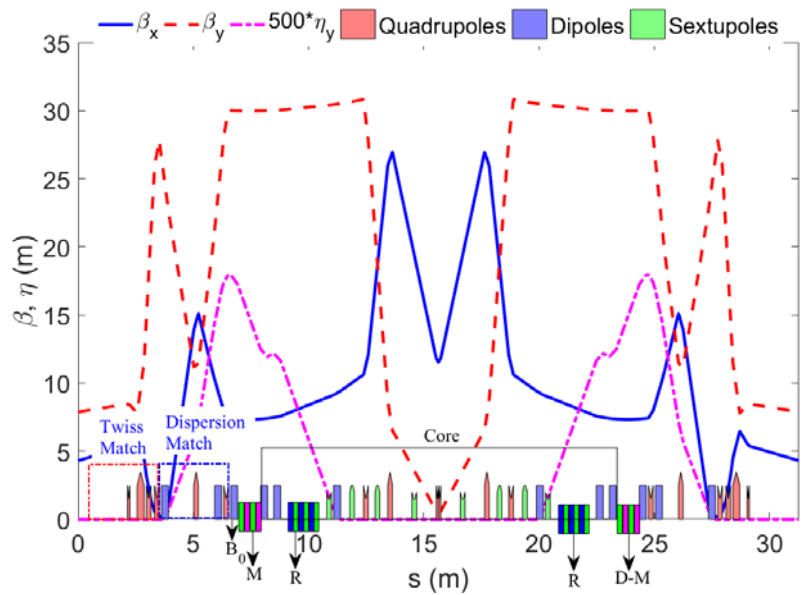

Figure 6. Beam optics for bypass section.

* fengchao@zjlab.org.cn

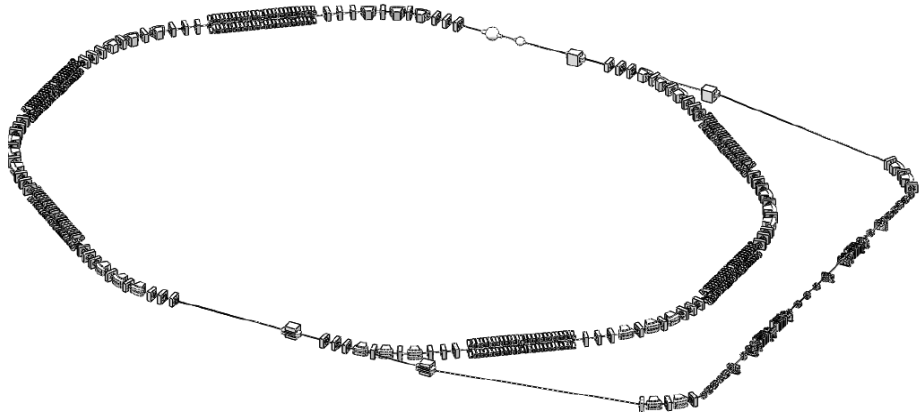

Figure 7. Sketch of damping ring and bypass beam line

The storage ring is filled by several bunch trains. The bunch train is kicked out to the bypass beam line successively for radiation. This kind of bunch train filling pattern can reduce the technical challenges of the kicker system, for which the repetition rate of the kicker will be reduced and the pulse width will be increased comparing to bunch-by-bunch kick out. With this pattern, the radiation will be in the burst mode.

Three-dimensional numerical simulations have been applied to show the possible performance of the proposed ring. Main parameters employed in the 3D simulations are given in TABLE II. The laser-electron beam interaction in the modulator induces an energy modulation amplitude of about 0.6 times of the initial energy spread (with IBS effect). The bunching factor distribution before entering the radiator (R) is shown in Fig. 8, where one can find that the bunching factor at $19^{th}$ harmonic (13.55 nm) is about 9%.

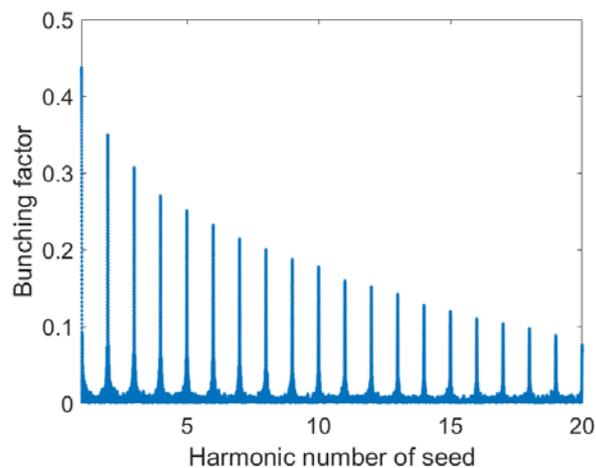

Figure 8. Bunching factor for ADM.

Fig.9. shows the residual energy modulation of the electron beam after passing through the whole bypass line. It can be seen that the residual energy modulation is significantly reduced after the optimization of sextupole magnets. The vertical emittance increases by 6.89% (RMS) and the energy spread increases by 0.016% of a

* fengchao@zjlab.org.cn

single pass. As the horizontal emittance is large and in an irrelevant plane, the emittance growth in horizontal plane is negligible.

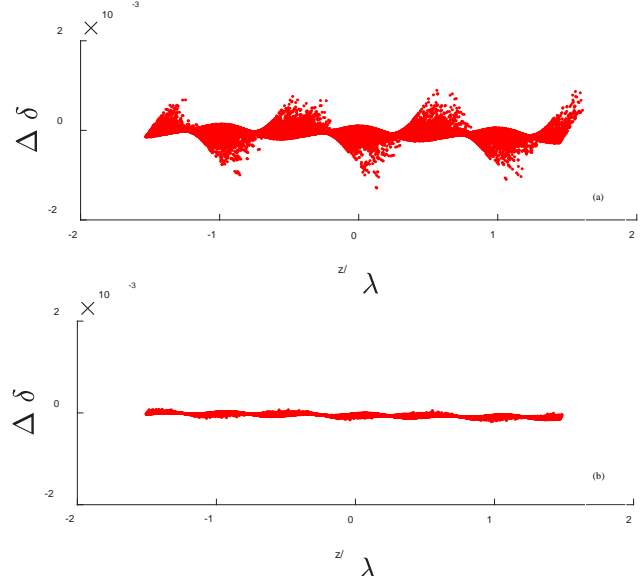

Figure 9. Residual energy modulation before (a) and after (b) sextupoles optimization.

The electron beam normally circulates in the storage ring. In a short repetition time, the electron beam will be kicked out to the bypass beam line again to interact with the laser and emit EUV radiation. The growth of the vertical emittance and the energy spread per turn is sufficiently small after demodulation, which results an unclear damping process when including quantum exciting effect. Therefore, we analyze the repetition rate according to the theoretical formula. The growth of vertical emittance and energy spread will be damped in the storage ring according to the following two formulas [7]:

$$\varepsilon_s(t) = \varepsilon_{s0} e^{-\frac{2t}{\tau_s}}, \tag{3}$$

$$\varepsilon_y(t) = \varepsilon_{y0} e^{-\frac{2t}{\tau_y}}, \tag{4}$$

where $\varepsilon_s$, $\varepsilon_y$ are the longitudinal and vertical emittances, $\varepsilon_{s0}$, $\varepsilon_{y0}$ is the bananced longitudinal and vertical emittances, $\tau_s$, $\tau_y$ are the longitudinal and vertical damping time.

The energy spread growth can be damped down in one turn. As the vertical emittance $\varepsilon_{y0}$ is very small which contributes very limited nonlinear effect on the isochronous beam line, the imperfect demodulation is majorly from the longitudinal drift caused by the longitudinal and the horizontal emittances via $T_{566}$, $T_{511}$, $T_{522}$ and $T_{512}$ terms. As the energy spread and the horizontal emittance are almost unchanged after demodulation, the growth of the vertical emittance is approximate an absolute value, which is about 0.64 pm·rad. The vertical emittance growth can be damped down in 95 turns. Therefore, the repetition rate of a single bunch is 20 kHz. Assuming the bunch number is 190, the repetition rate of EUV radiation for a single pulse mode is about 3.8MHz.

* fengchao@zjlab.org.cn

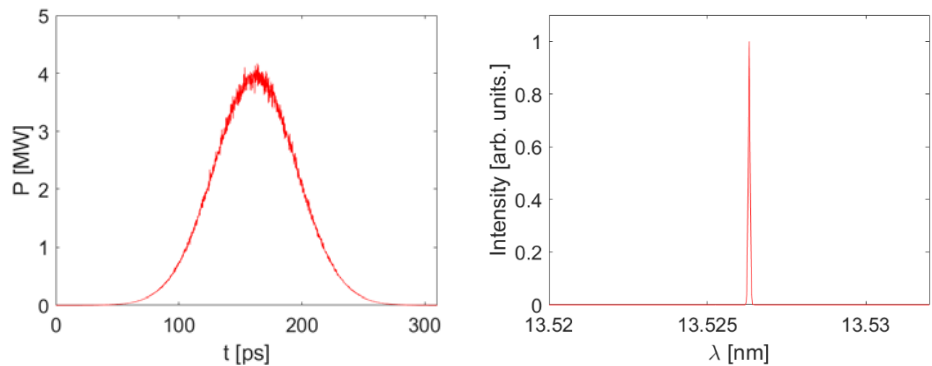

Figure 10. Output radiation pulse and the corresponding single-shot spectrum

The longitudinal profile and the corresponding spectrum of a single EUV radiation pulse simulated by Genesis [25] are shown in Fig.10. The single pulse energy is about 332 μJ, which is produced by a 3.5m long undulator with period length 2.5cm. With a repetition rate of 3.8MHz, the average power is calculated to be about 1.26kW. There are 2 undulators in the beam line as indicated in Fig.6 with a canted angle of 19mrad in vertical plane. The total output average power of the proposed storage ring reaches 2.52kW.

## 4. Discussion

The instabilities should be carefully studied for high current operation of the ring, however, 15.8mA/bunch current is not an aggressive number. IBS effect has already estimated in section II. Other issues will not be discussed in this paper in detail. A rough estimation is that multi-bunch instability with an order of magnitude higher current will be damped by an order of magnitude lower damping time comparing to an ordinary storage ring. The vacuum pipe should be carefully designed to avoid wake field energy loss at the small steps to avoid beam pipe been heated.

The radiation power produced per straight section from SWs is about 330 kW which is great but manageable. The radiation divergence from the wiggler is 7.8mrad and 0.3mrad in H/V planes respectively. At the end of the wiggler, the diameters of the spot are 67.1mm (H) and 2.7mm (V). The size of SW beam pipe can be larger than those values to avoid a major energy dissipate on the SWs beam pipe. The radiation power from SWs can be absorbed by a specially designed high-power absorber in the following arc. Such kind of absorber (256 kW) has been designed for the ILC damping ring[26].

RF system is a tough job for this high current storage ring which should provide 2100kW RF power to the beam. Due to the low accelerating voltage and high beam loading operation parameters, the normal conducting technology would be adopted. The RF input coupler and the HOM coupler/absorber should be the key components of the main cavities.

2.5 kW EUV radiation has been gotten with a 3A average beam current. The energy transfer efficiency from the beam to the EUV radiation is more than 0.1%. The power is mainly consumed by SWs. The EUV radiation from SWs can be connected if it gets value which about 19.8W for each.

* fengchao@zjlab.org.cn

The damping ring itself gets outstanding performances with large DA and LMA. We have also tried the case with lower beam energy of 600MeV, the nature emittance is 0.152nm·rad which means the normalized beam emittance is only 0.178μm·rad. This damping ring can be a competitive candidate for the injectors of colliders or free electron lasers.

ACKNOWLEDGMENTS

This work is supported by the National Natural Science Foundation of China (No. 11975298, No. 11975300) and Shanghai Science and Technology Committee Rising-Star Program (20QA1410100). We thank Ya Zhu and Shengwang Xiang for figure preparation.

* fengchao@zjlab.org.cn

* fengchao@zjlab.org.cn